\documentclass[aps,prb,twocolumn,superscriptaddress,showkeys,citeautoscript]{revtex4-1}

\usepackage[version=4]{mhchem} 
\usepackage[hidelinks,colorlinks=true,allcolors=blue]{hyperref}
\usepackage{subfigure}
\usepackage{graphicx}
\graphicspath{{./figures/}}
\usepackage{amsmath} %
\usepackage{color} 
\usepackage{float} 
\setcitestyle{super}
\usepackage{siunitx}

\newcommand{\etal}{et~al.}

\newcommand{\kapa}{$\kappa_{\ell}$~}
\newcommand{\bsi}{$\textup{B}_{\textup{Si}}^{(-1)}$}

\newcommand{\psipd}{$\textup{P}_{\textup{Si}}^{(+1)}$}

\begin{document}


\title{Combined treatment of phonon scattering by electrons and point defects explains the thermal conductivity reduction in highly-doped Si}


\author{Bonny Dongre}
\email[]{bonny.dongre@tuwien.ac.at}
\author{Jes\'us Carrete}
\affiliation{Institute of Materials Chemistry, TU Wien, A-1060 Vienna,  Austria}

\author{Shihao Wen}
\author{Jinlong Ma}
\author{Wu Li}
\affiliation{Institute for Advanced Study, Shenzhen University, Shenzhen 518060, China}

\author{Natalio Mingo}
\affiliation{LITEN, CEA-Grenoble, 17 rue des Martyrs, 38054 Grenoble Cedex 9, France}

\author{Georg K. H. Madsen}
\affiliation{Institute of Materials Chemistry, TU Wien, A-1060 Vienna,  Austria}


\date{\today}

\begin{abstract}
The mechanisms causing the reduction in lattice thermal conductivity in highly P- and B-doped Si are looked into in detail. Scattering rates of phonons by point defects, as well as by electrons, are calculated from first principles. Lattice thermal conductivities are calculated considering these scattering mechanisms both individually and together. It is found that at low carrier concentrations and temperatures phonon scattering by electrons is dominant and can reproduce the experimental thermal conductivity reduction. However, at higher doping concentrations the scattering rates of phonons by point defects dominate the ones by electrons except for the lowest phonon frequencies. Consequently, phonon scattering by point defects contributes substantially to the thermal conductivity reduction in Si at defect concentrations above $10^{19}$~cm$^{-3}$ even at room temperature. Only when, phonon scattering by both point defects and electrons are taken into account, excellent agreement is obtained with the experimental values at all temperatures. 
\end{abstract}


\maketitle

\section{\label{sec:intro} Introduction}
With the characteristic lengths of nanoscale devices approaching the mean free paths of the heat-carrying phonons,\cite{cahill2014nanoscale} the need for a detailed and predictive understanding of thermal conductivity is more stark today than ever. In the last few decades, the usage of highly doped ($10^{19}-10^{21}~\textup{cm}^{-3}$) semiconductor materials in electronics and energy devices has become prevalent, serving the purpose of achieving enhanced functional properties.~\cite{fistul2012heavily} For these systems, predictive models are especially important  because the increased phonon scattering due to point defects and additional charge carriers results in a substantial drop in thermal conductivity. 

Owing to its high abundance, non-toxicity, and ease of dopability, Si continues to be the linchpin of the semiconductor industry. Highly P- and B-doped Si are routinely used as source/drain materials in transistors to avoid unwanted Schottky junctions.\cite{sze2006physics} Furthermore, highly-doped Si has also found usage in photovoltaics~\cite{ehsani1997optical}, microelectromechanical systems~\cite{ng2015temperature}, and  microelectronics~\cite{beyers1987titanium, vandort1994a}, to cite a few applications. In  thermoelectric applications the advantage of using highly-doped Si is twofold:\cite{neophytou2013simultaneous,ohishi2015thermoelectric, zhu2016the} the thermoelectric figure of merit is, on the one hand, proportional to the electronic power factor, which increases with increasing carrier concentration, and, on the other, inversely proportional to the thermal conductivity. 

The lattice contribution to the thermal conductivity ($\kappa_{\ell}$)  dominates in Si. Phonon scattering by point defects (PDPS) and by electrons (EPS), were identified as the two main contributors to the \kapa reduction in highly-doped Si.~\cite{abels1963lattice, ohishi2015thermoelectric, liao2015significant, zhu2016the} 
 Zhu~\etal~\cite{zhu2016the} reported a ${\approx}36\%$ reduction in \kapa in fine-grained, highly P-doped Si, and asserted that EPS is the major contributor to the \kapa reduction. 
 In contrast, Ohishi~\etal~\cite{ohishi2015thermoelectric} attributed the \kapa reduction in single-crystal, highly P- and B-doped Si solely to intrinsic anharmonic phonon-phonon scattering and PDPS. 
 
The aforementioned disagreement highlights the problem of separating scattering mechanisms when employing fitted models based on strongly-simplified assumptions about the underlying phonon band structures and scattering mechanisms. Important progress towards a more predictive treatment of \kapa in doped Si was made recently by Liao~\etal~\cite{liao2015significant}, who performed an \textit{ab-initio} study of $n$- and $p$-doped Si and showed that, EPS at a carrier concentration of $p\approx 10^{21}$~cm$^{-3}$ can result in a ${\approx}45\%$ reduction in \kapa at room temperature. The calculations reproduce how \kapa is lower in $p$-doped samples than in $n$-doped ones, in agreement with the experiments.\cite{slack1964thermal,asheghi2002thermal} However, they do not capture the magnitude of the reduction observed in B-doped $p$-type single-crystal Si, which at a doping level of $5\times10^{21}$~cm$^{-3}$ amounts to more than 70\%.\cite{slack1964thermal}

In the present work, we investigate the precise mechanisms responsible for the \kapa reduction observed in highly-doped Si. We calculate \kapa by employing the Boltzmann transport equation (BTE) for phonons, using only inputs in the form of interatomic force constants (IFCs) and electron-phonon coupling (EPC) matrix elements obtained from density functional theory. We extend the earlier work on EPS\cite{liao2015significant} and include also the PDPS from first principles. At high defect concentrations, we find that the PDPS rates dominate the EPS rates at all frequencies except the lowest ones and contribute substantially to the \kapa reduction at all temperatures. On the other hand, EPS dominates at low defect concentrations due to a fundamentally different frequency and concentration dependence. As a result, a correct quantitative prediction of the \kapa dependence on defect concentration and temperature is obtained only when both EPS and PDPS are taken into account.

\section{\label{sec:method}Methodology}

Within the relaxation-time approximation, the lattice thermal conductivity tensor can be expressed as~\cite{li2014shengbte, dongre2018abinitio}
\begin{equation}
\kappa^{\alpha \beta}_{\ell} = \dfrac{1}{k_B T^2 V_{\textup{uc}}} \sum_{i\mathbf{q}} n^0_{i\mathbf{q}}(n^0_{i\mathbf{q}} + 1) (\hbar \omega_{i\mathbf{q}})^2 v_{i\mathbf{q}}^{\alpha} v_{i\mathbf{q}}^{\beta} \tau_{i\mathbf{q}}^{0}, 
\label{eq:kappa}
\end{equation}
where $\alpha$ and $\beta$ run over the Cartesian axes, $k_B$ is the Boltzmann constant, $V_{\textup{uc}}$ is the unit cell volume, $n^0_{i\mathbf{q}}$, $v_{i\mathbf{q}}^{\alpha}$,and $\omega_{i\mathbf{q}}$ are the Bose-Einstein occupancy, the group velocity, and the angular frequency of a phonon mode with wave-vector $\mathbf{q}$ and branch index $i$, respectively. $\tau^0_{i\mathbf{q}}$ represents the relaxation time of mode ${i\mathbf{q}}$ and is obtained as:
\begin{equation}
\dfrac{1}{\tau^{0}_{i\mathbf{q}}} = \dfrac{1}{\tau^{\textup{3ph}}_{i\mathbf{q}}} + \dfrac{1}{\tau^{\textup{iso}}_{i\mathbf{q}}} + \dfrac{1}{\tau^{\textup{pd}}_{i\mathbf{q}}} + \dfrac{1}{\tau^{\textup{ep}}_{i\mathbf{q}}},
\label{eq:tau}
\end{equation}
where the lifetime of a phonon of mode $i\mathbf{q}$ as limited by the scattering caused by: the three-phonon processes is given by $\tau^{\textup{3ph}}_{i\mathbf{q}}$, the mass disorder due to isotopes by $\tau^{\textup{iso}}_{i\mathbf{q}}$, the point defects by $\tau^{\textup{pd}}_{i\mathbf{q}}$, and the electrons by $\tau^{\textup{ep}}_{i\mathbf{q}}$. The expressions for the phonon scattering by three-phonon processes and isotopes  can be found in Ref.~\citenum{li2014shengbte}. 

\subsection{\label{subsec:pointdefscat} Point-defect phonon scattering}
The point defect phonon scattering rates, $1/\tau^{\textup{pd}}_{i\mathbf{q}}$, can be calculated as~\cite{mingo2010cluster}:
\begin{equation}
  \dfrac{1}{\tau_{i\mathbf{q}}^{\textup{pd}}}  = -n_\mathrm{def} V_\mathrm{uc} \dfrac{1}{\omega_{i\mathbf{q}}}\textup{Im}\left \{   {\mathbf{e}}_{i\mathbf{q}}^{\dagger} \mathbf{T}(\omega^2)   {\mathbf{e}}_{i\mathbf{q}} \right   \},
  \label{eq:scatteringoptical}
\end{equation}
where $n_\textup{def}$ is the volumetric concentration of the point defects  and $\mathbf{e}_{i\mathbf{q}}$ represents an incoming phonon mode. $\mathbf{T}$ is the matrix that relates the Green's function of the perturbed lattice to that of the unperturbed lattice. It can be represented in terms of the retarded Green's functions of the unperturbed host lattice, $\mathbf{g}^+(\omega^2)$, and the perturbation $\mathbf{V}$ as~\cite{economou1983green}:
\begin{equation}
\mathbf{T} = (1 -   \mathbf{V} \mathbf{g}^+)^{-1} \mathbf{V}.
\label{eq:T}
\end{equation}
The matrix element of $\mathbf{g}^+$ projected on the atom pairs $l\eta$ and $l'\eta'$ is given as~\cite{economou1983green}:
\begin{equation}
{\mathbf{g}^+}_{l\eta, l'\eta'}(\omega^2)   =  \lim_{\epsilon \to 0^+} \sum_{i\mathbf{q}} \dfrac{\mathbf{e}_{i\mathbf{q}}(l\eta) \mathbf{e}^{\dagger}_{i\mathbf{q}}(l'\eta')}{\omega^2  + i\epsilon - \omega^2_{i\mathbf{q}}},
\end{equation}
where, $\mathbf{e}_{i\mathbf{q}}(l\eta)$ is the eigenvector of mode $i\mathbf{q}$ projected on the $\eta$-th atom in the $l$-th unit cell. Moreover,
\begin{equation}
\mathbf{V} = \mathbf{V}_\textup{M} + \mathbf{V}_\textup{K},
\end{equation}
where $\mathbf{V}_\textup{M}$ and $\mathbf{V}_\textup{K}$ are the mass and IFC perturbation matrices, respectively. Their matrix elements are given by:
\begin{align}
V_{\textup{M}; l\eta,l\eta} &= \frac{M_{l\eta,l\eta} - M_{0;l\eta,l\eta}}{M_{0;l\eta,l\eta}}\omega^2, \text{and} \nonumber \\ 
V_{\textup{K}; l\eta, l'\eta'}^{\alpha \beta} &= - \frac{\Phi^{\alpha \beta}_{l\eta, l'\eta'} - \Phi_{0;l\eta, l'\eta'}^{\alpha \beta}}{\sqrt{M_{l\eta} M_{l'\eta'}}},
\end{align}
where $M$ is the mass of the defect atom which replaces a host atom of mass $M_0$ at site $l\eta$. $\Phi$ and $\Phi_0$ are the IFC matrices of the defect-laden and host systems, respectively.

\subsection{\label{subsec:elphscat} electron-phonon scattering}
The electron-phonon scattering rates can be expressed as~\cite{liao2015significant, ziman2001electrons}:
\begin{equation}
    \begin{split}
  \dfrac{1}{\tau_{i\mathbf{q}}^{\textup{ep}}}  &= \dfrac{2\pi}{\hbar} \sum_{mn,\mathbf{k}} \left | {g}_{mn}^{i} (\mathbf{k}, \mathbf{q})\right |^2 (f_{m\mathbf{k}+ \mathbf{q}} - f_{n\mathbf{k}} )  \\  
  &\quad\times \delta (\varepsilon_{n\mathbf{k}} - \varepsilon_{m\mathbf{k} + \mathbf{q}} - \hbar \omega_{i\mathbf{q}}),
\end{split}
 \label{eq:epscattering}
\end{equation}
where ${g}_{mn}^{i} (\mathbf{k} ,\mathbf{q})$ is the EPC matrix element of an interaction process involving a given phonon $i\mathbf{q}$ and two charge carriers with band indices $m$ and $n$ and wave-vectors $\mathbf{k}$ and $\mathbf{k} + \mathbf{q}$, respectively. $f_{n\mathbf{k}}$ is the Fermi-Dirac distribution function and $\varepsilon _{n\mathbf{k}}$ is the eigenenergy of an electron state $n\mathbf{k}$.
In non-spin-orbital-coupling or non-magnetic calculations, the formula above must be multiplied by a factor of two to take the electron spin degeneracy into account. 

The EPC matrix element can be computed within density functional perturbation theory as~\cite{li2015electrical}:
\begin{equation}
{g}_{mn}^{i} (\mathbf{k}, \mathbf{q})=\sqrt{\dfrac{\hbar}{2  \omega_{i\mathbf{q}} } } \sum_{\eta\alpha} \dfrac{{e}_{i\mathbf{q}}^{\alpha}(0\eta)}{\sqrt{M_\eta}}\bigg\langle{m\mathbf{k}+\mathbf{q}\left|\dfrac{\partial V_\text{KS}(\mathbf{r})}{\partial {u}_{i\mathbf{q}}^{\alpha}(0\eta)}\right|n\mathbf{k} }\bigg\rangle
\end{equation}
where $M_{\eta}$ is the atomic mass of the $\eta$-th atom in the unit cell, ${\alpha}$ is the Cartesian direction, and $\partial V_\text{KS}(\mathbf{r})/\partial {u}_{i\mathbf{q}}^{\alpha}(0\eta)$ is the perturbation of the Kohn-Sham potential with respect to the displacement ${u}_{i\mathbf{q}}^{\alpha}(0\eta)$.

\section{\label{sec:compdetail} Computational Details}
For calculating the PDPS rates, the total energy as well as the force calculations are carried out using the projector-augmented-wave method~\cite{blochl1994projector} as implemented in the VASP code~\cite{kresse1999from}, within the local density approximation (LDA)~\cite{Kohn_PR65,perdew1981self} to the exchange and correlation energy. For completeness and comparison to LDA we also calculate the PDPS rates with the Perdew-Burke-Ernzerhof (PBE) exchange-correlation
functional~\cite{perdew1996generalized}.
The volume of the unit cell is relaxed until the energy is converged within $10^{-8}$~eV. The 2$^\mathrm{nd}$- and 3$^\mathrm{rd}$-order IFCs are extracted using $5\times5\times5$ supercell of the rhombohedral primitive cell containing 250 atoms using just the $\Gamma$-point. For the 2$^\mathrm{nd}$-order IFC calculations we use the Phonopy~\cite{togo2015first} software package and for the 3$^\mathrm{rd}$-order IFCs we use \texttt{thirdorder.py} from the ShengBTE package~\cite{li2014shengbte}. The same supercell size was used to calculate the IFC for the defect-laden structures. To compute the PDPS rates, the Green’s functions are calculated on a $38\times38\times38$ q-point mesh using the linear tetrahedron method~\cite{lambin1984computation} for integration over the Brillouin zone.  The scattering rates are then calculated on $35\times35\times35$ q-point mesh.

For the calculation of EPS rates, the EPC matrix elements are first computed on coarse grids and then interpolated to dense grids with the Wannier function interpolation method\cite{giustino2007electron, marzari2012maximally}. The interpolations are performed using Quantum Espresso \cite{giannozzi2009quantum} and the built-in EPW package\cite{noffsinger2010epw} with norm-conserving pseudopotentials. Likewise, both the LDA and PBE exchange and correlation functionals are considered. The initial $\mathbf{k}$ and $\mathbf{q}$ grids are both $6\times6\times6$, which are interpolated to $35\times35\times35$ meshes needed for the thermal conductivity calculations. The energy conservation $\delta$-function is treated by Gaussian function with self-adaptive broadening parameters.\cite{li2015electrical} 

Finally, the bulk thermal conductivity is also calculated using the $35\times35\times35$ q-point mesh with the \textsc{almaBTE}~\cite{carrete2017almabte} package. Due to the dense $\mathbf{q}$-mesh the thermal conductivity is converged down to 40~K.


\section{\label{sec:results}Results and discussion}

\begin{figure*}
 \centering
 \subfigure[P-doped Si]{ \label{} \includegraphics[width=.45\textwidth]{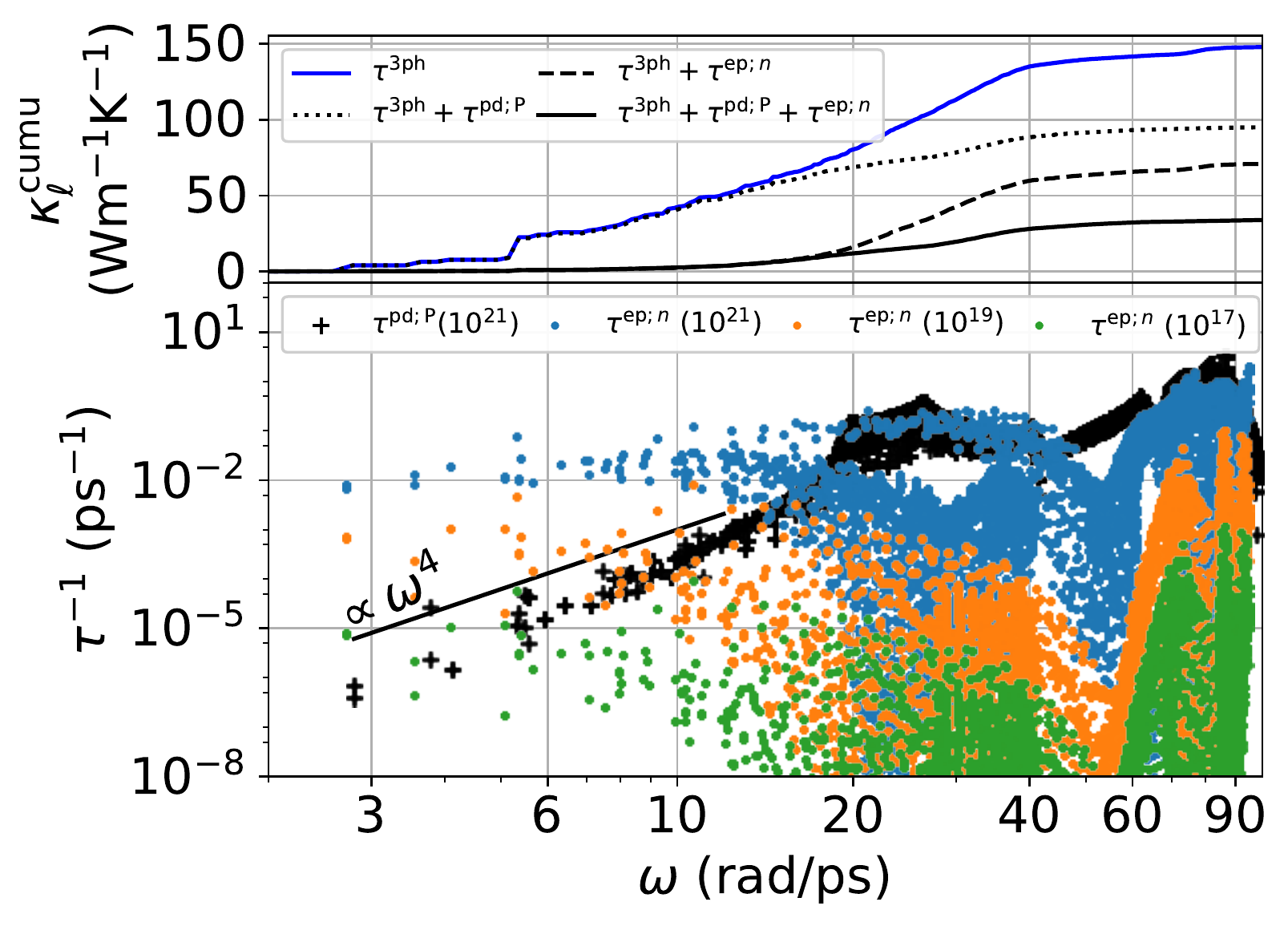}}
 \subfigure[B-doped Si]{\label{} \includegraphics[width=.45\textwidth]{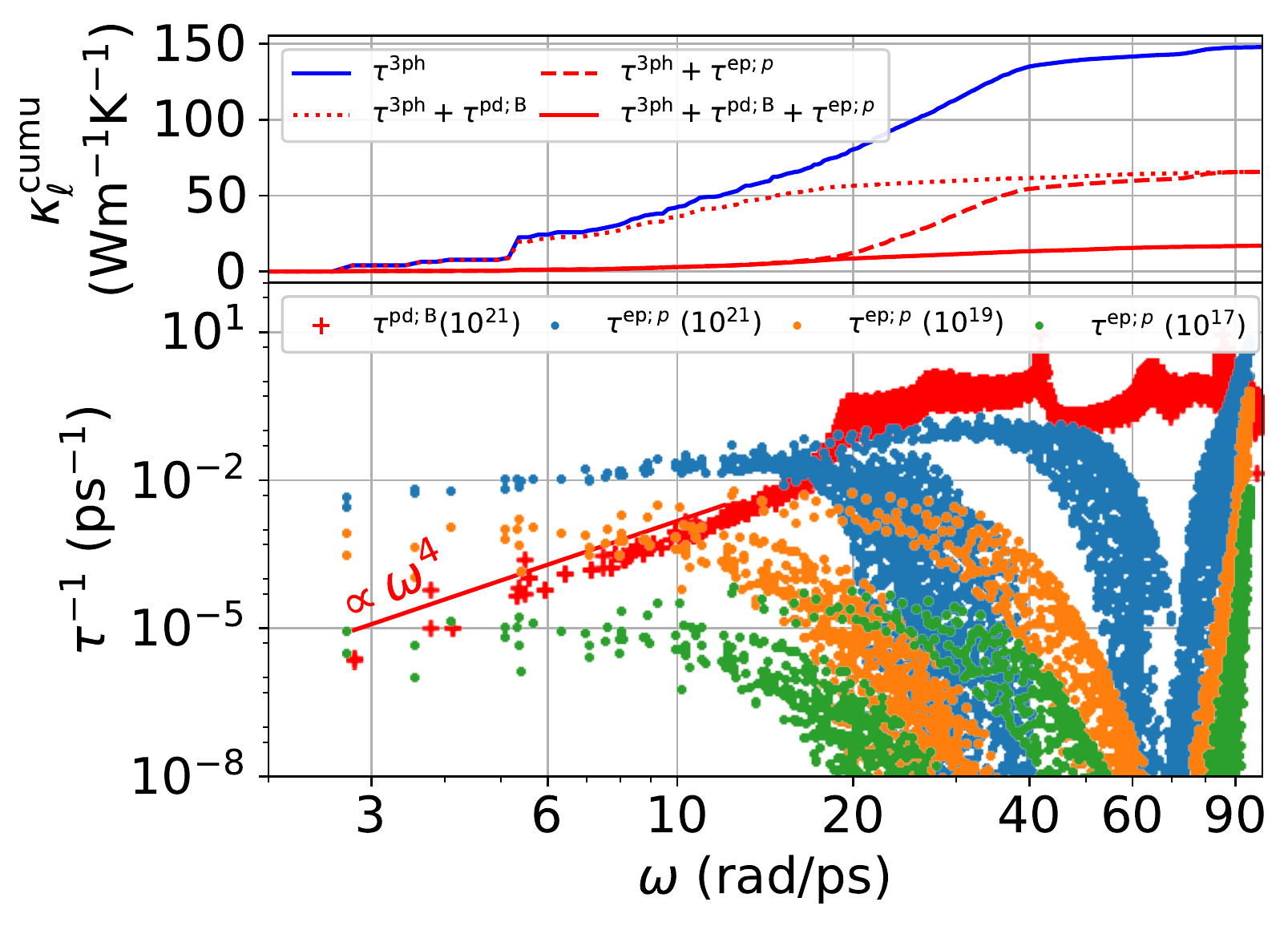} }
 \caption{EPS (dots) and PDPS (+ signs) rates for highly-doped Si. PDPS rates are shown at a doping concentration of $10^{21}~\textup{cm}^{-3}$ and the EPS rates are shown for three different carrier concentrations at 300~K. The top panels in the figures show the cumulative \kapa at 300~K and a defect concentration of $10^{21}~\textup{cm}^{-3}$. In all the \kapa calculations reported in these  and later figures, the phonon scattering caused by the mass disorder due to isotopes, $\tau^{\textup{iso}}$, is also considered by default.} 
\label{fig:pdelescatrate}
\end{figure*}
\begin{figure}
 \includegraphics[scale = .54]{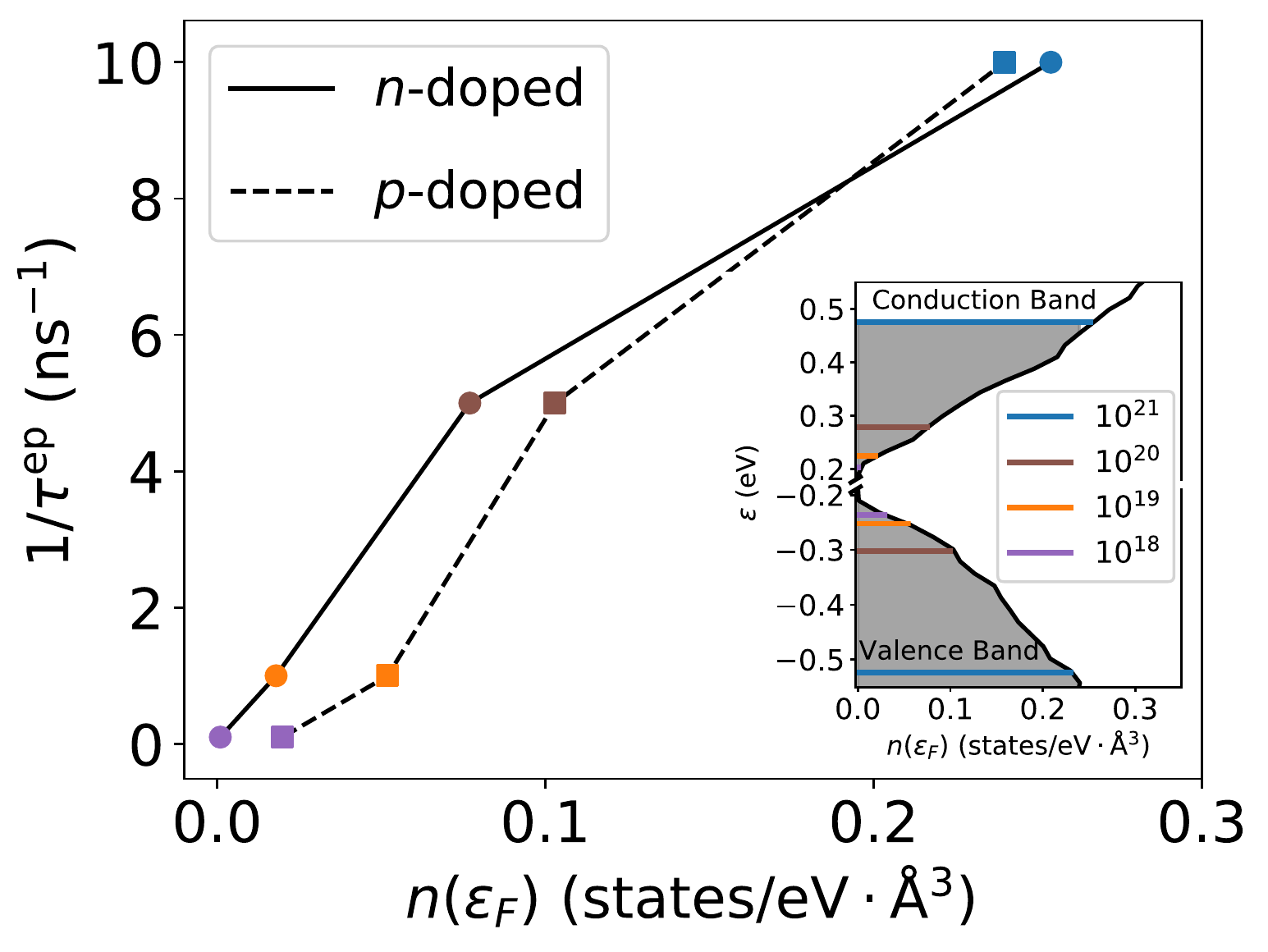}%
 \caption{\label{fig:tau_vs_dos} The low-frequency EPS rates for different concentrations as a function of the electronic density of states. The marker colors correspond to the doping levels in the inset. }
\end{figure}

Figure~\ref{fig:pdelescatrate} shows a comparison between the PDPS and EPS rates for both \textit{n}- and \textit{p}-doped Si, in the bottom panels. The PDPS depends on the actual defect type. We have studied the substitutional \bsi and \psipd defects, which are the most common doping elements and also those studied experimentally. The PDPS has a trivial dependence on the defect concentration [Eq.~\eqref{eq:scatteringoptical}] and is shown for only one defect concentration (10$^{21}$~cm$^{-3}$). In a rigid band approximation, the EPS is dependent only on the carrier concentration through the distribution function in Eq.~\eqref{eq:epscattering}. The EPS is shown for three different carrier concentrations in Figure~\ref{fig:pdelescatrate} and the corresponding chemical potentials are shown in Fig.~\ref{fig:tau_vs_dos} (inset). Our LDA EPS rates compare well with the PBE rates reported by Liao \etal \cite{liao2015significant}. The LDA rates are slightly higher than the PBE ones which results in a correspondingly higher \kapa reduction due to EPS in LDA as compared to PBE. However, this does not have any major effect on the interpretation of our results. 

If we first consider the low frequency ($\omega<12$~rad/ps) behavior, the PDPS rates exhibit a simple Rayleigh $\omega^4$ behavior and the EPS rates for a given carrier concentration are close to being independent of $\omega$. Plotting the EPS rates as a function of the electronic density of states (DOS) at the electron chemical potential corresponding to a given doping shows 
 an almost linear dependence (Fig.~\ref{fig:tau_vs_dos}). The EPS rates thus behave in accordance with a simple $\tau^{-1}\propto n(\varepsilon)$ model\cite{Xu_PRL14} at low frequencies. With a $n(\varepsilon)\propto\varepsilon^{1/2}$ behavior of the electronic DOS, the low-frequency EPS rates will scale approximately as $n_{\mathrm{def}}^{1/3}$ with the carrier/defect concentration as opposed to the linear scaling of the PDPS rates evident from Eq.~\eqref{eq:scatteringoptical}. These simple relations are in accordance with the expectations that EPS will dominate over PDPS at low temperatures and defect concentrations while PDPS will become increasingly important at high defect concentrations.
 
At the same time, it is clear from Fig.~\ref{fig:pdelescatrate} that for $\omega>12$~rad/ps the calculated rates deviate substantially from the aforementioned simple relations. The cumulative \kapa plots in the respective top panels in Fig.~\ref{fig:pdelescatrate} (blue curves) illustrate that modes with $\omega>12$~rad/ps carry about two-thirds of the heat at 300~K. Simply extrapolating the low frequency behavior would lead to a strong overestimation of the scattering.
This is especially so for the EPS rates where a simple extrapolation would result in a strong overestimation of the predicted \kapa suppression. The cumulative plots in the top panels of Fig.~\ref{fig:pdelescatrate} also show for PDPS (dotted lines), at a large defect concentration, that the contribution to \kapa for frequencies higher than 20~rads/ps is only ${\approx}15\%$ in case of \bsi defect and ${\approx}40\%$ in case of \psipd. In contrast, the EPS causes a majority reduction in \kapa by frequencies below 20~rads/ps, for both P- and B-doping. 

\begin{figure}[t]
 \includegraphics[scale = .55]{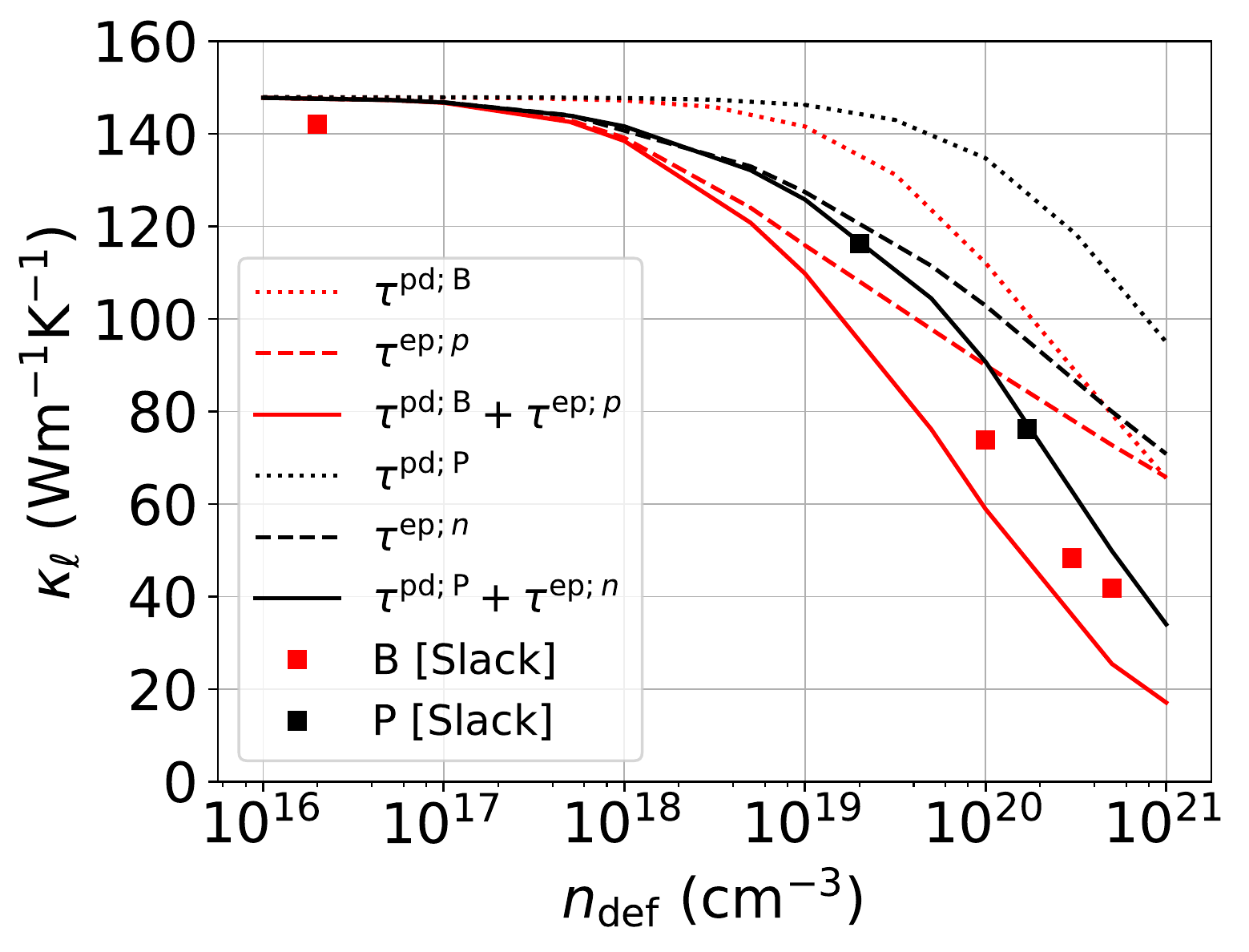}%
 \caption{\label{fig:pselboth} Comparison of the reduction in \kapa caused by EPS and PDPS individually and combined together vs increasing doping concentration. The filled squares are from the experimental data in Ref.~\citenum{slack1964thermal}. }
\end{figure}

Next, in Fig.~\ref{fig:pselboth}, we look into the individual contributions from the EPS (dashed lines) and PDPS (dotted lines) to the room temperature thermal conductivity reduction for increasing doping concentrations and compare them to the experimental \kapa data from Slack.\cite{slack1964thermal} In accordance with the analysis of the scattering rates, EPS dominates the reduction of \kapa at low carrier concentrations. In the case of $n$-doped Si, EPS alone is almost enough to explain the experimental point at $2\times 10^{19}~\mathrm{cm}^{-3}$. However, at higher carrier concentrations EPS alone clearly underestimates both the absolute reduction of \kapa and the trend. Interestingly, PDPS captures the trend correctly for large defect concentrations, but also underestimates the absolute \kapa reduction. At a doping concentration of $10^{21}~\mathrm{cm}^{-3}$, the EPS and PDPS contribute almost equally in \kapa reduction for B doping. Even though both EPS and PDPS contribute substantially to \kapa reduction in Si, neither can explain the absolute reduction in \kapa on its own. Only when both are taken into consideration in Eq.~\eqref{eq:tau} is the experimentally observed reduction of $\kappa_{\ell}$ reproduced. This is shown by the black and red solid lines in Fig.~\ref{fig:pselboth}. Apart from a slight underestimation of \kapa in case of B-doping, the calculated value of \kapa considering both EPS and PDPS agree very well with the experimental values available for concentrations ${\sim}10^{19}-10^{21}$ cm$^{-3}$ both in value and trend, Fig.~\ref{fig:pselboth}.  

\begin{figure}[t]
 \includegraphics[scale = .55]{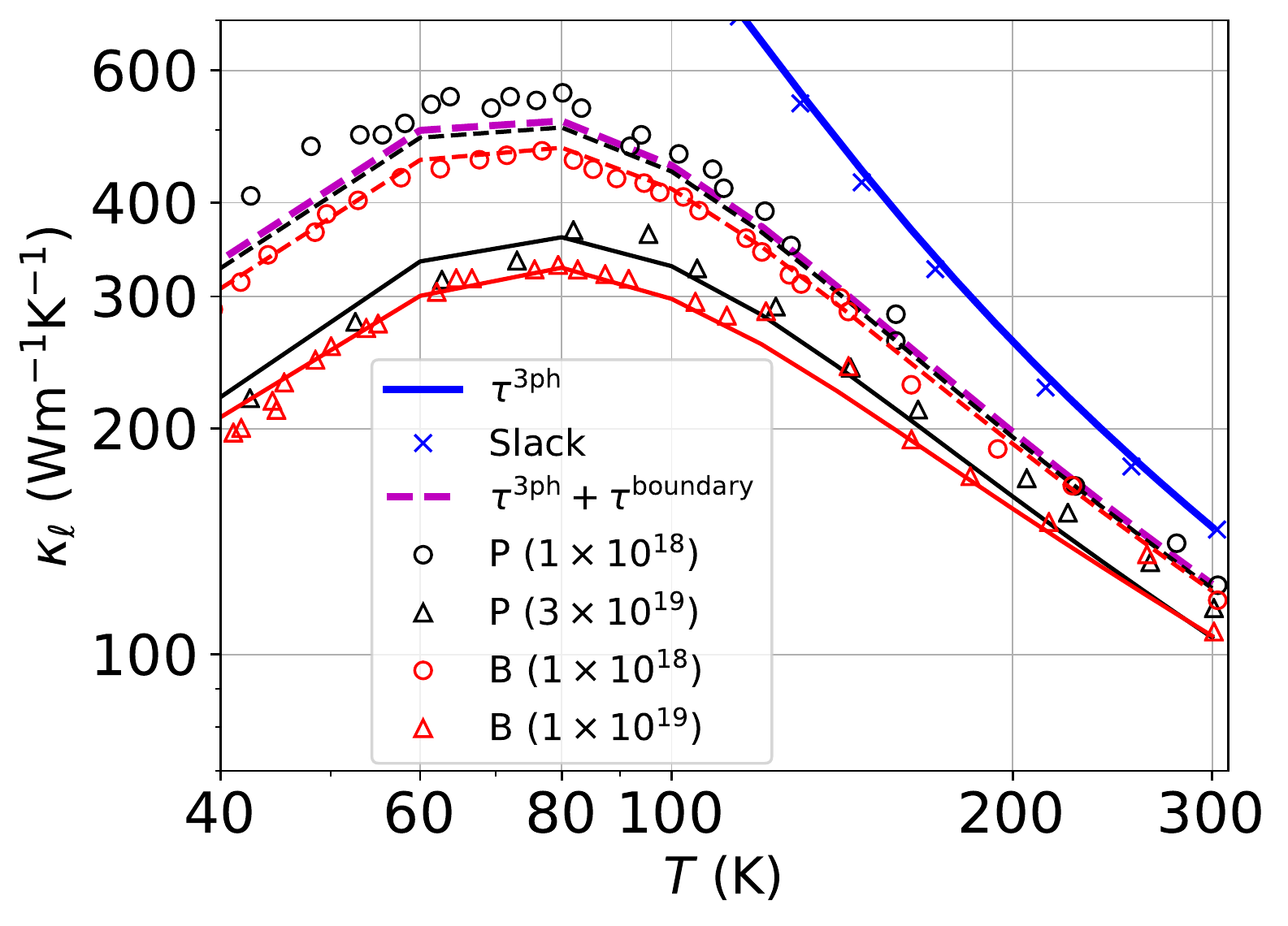}%
 \caption{\label{fig:kappavst} \kapa vs temperature curves for the B(red)- and P(black)-doped Si considering the three-phonon, PDPS, EPS, and boundary scattering (at \SI{10}{\micro\meter}). The open triangles and circles are the experimental data obtained from Ref.~\citenum{asheghi2002thermal}. The solid and dashed (black and red) lines correspond to the calculations done at experimental carrier concentrations.}
\end{figure}

Besides the work of Slack~\etal~\cite{slack1964thermal} used for Fig.~\ref{fig:pselboth}, there is a general lack of systematic experimental data on the thermal conductivity of single-crystal Si with varying doping concentrations. 
In order to gain further confidence in our results, we compare them to the more recent experimental data from Ref.~\citenum{asheghi2002thermal}, which were measured on single-crystal Si films. Even at low defect concentrations, $n_{\mathrm{def}}=10^{17}-10^{18}$~cm$^{-3}$, these samples exhibit a substantially lower \kapa than the bulk samples.\cite{asheghi2002thermal} However, a good agreement with the experimental curves can be obtained by adding a simple boundary scattering term, $1/\tau^B_{i\mathbf{q}}=|\mathbf{v}_{i\mathbf{q}}|/L$ with  $L=$ \SI{10}{\micro\meter}, to Eq.~\eqref{eq:tau} to emulate the effect of a film, as seen in Fig.~\ref{fig:kappavst} (purple line). Adding now the effect of the PDPS and EPS we calculate the variation of \kapa as a function of temperature for both B- and P-doped Si films, shown in Fig.~\ref{fig:kappavst}. For B-doping, when we include the PDPS and EPS along with the boundary scattering, we see that there is only a slight reduction from the purple line for the $10^{18}$ cm$^{-3}$ doping level (dashed red lines). Nevertheless, this results in an excellent agreement with the experimental data at that concentration (red circles), throughout the temperature range. Keeping the boundary scattering constant, when the doping concentration is increased to $10^{19}$ cm$^{-3}$, a large reduction in \kapa is observed (solid red line) which also agrees well with the experimental data (red circles). Similarly, for the P-doped calculations, we obtain an excellent agreement with the experiments except for a slight underestimation in \kapa at temperatures below 80 K for $10^{18}$~cm$^{-3}$ doped case. However, this is still under the uncertainties in experimental data. 

\begin{figure}[t]
 \includegraphics[scale = .55]{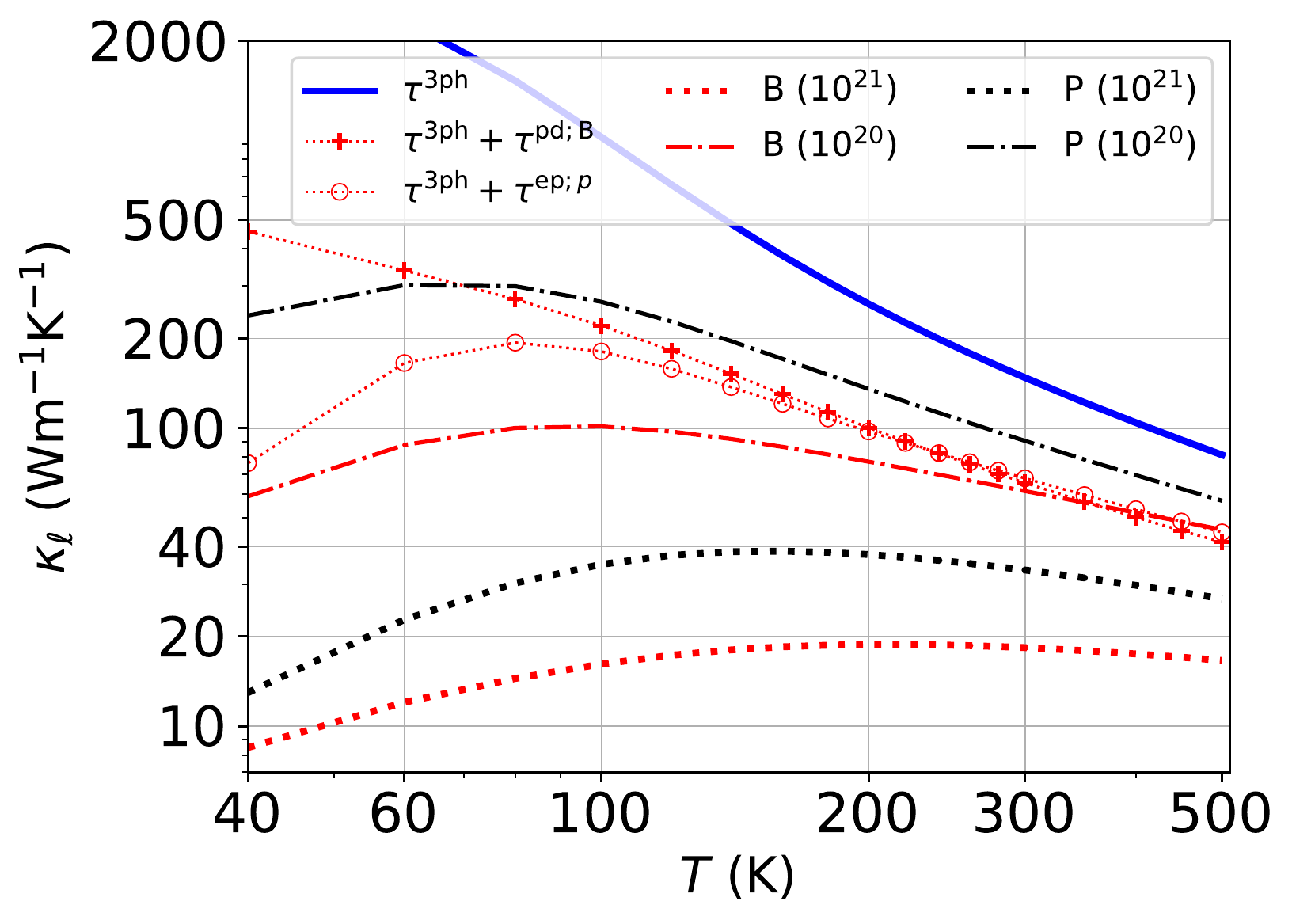}%
 \caption{\label{fig:kappavstb21} Prediction of \kapa for highly P- and B-doped Si. Also shown is the comparison of EPS and PDPS to \kapa reduction considered at a concentration of $10^{21}$ cm$^{-3}$.  }
\end{figure}

We then make predictions for the highly-doped cases ($10^{20}$ and $10^{21}$ cm$^{-3}$) as there is no experimental thermal conductivity data available for such high doping levels. These are shown in Fig.~\ref{fig:kappavstb21}. At room temperature, we find more than 60\% reduction as compared to the bulk value at $10^{20}$ cm$^{-3}$ doping level and 90\% at $10^{21}$ cm$^{-3}$ for B doping, and 40\% and 80\% for P doping, respectively. 
Fig.~\ref{fig:kappavstb21} also shows the individual contributions to the \kapa reduction by EPS and PDPS in the case of B doping at a doping concentration of $10^{21}$ cm$^{-3}$, revealing their characteristic effects on the thermal conductivity over the temperature range. The reduction in \kapa caused by PDPS overtakes the one by EPS at ${\approx}250$ K. This behavior highlights the importance of PDPS at high doping concentrations in Si and also emphasizes that, at temperatures higher than 300 K, PDPS is the dominant scattering mechanism besides the intrinsic anharmonic scattering. We have also performed calculations including an $L=$\SI{10}{\micro\meter} boundary scattering term, however at such high doping concentrations the EPS and PDPS dominate the boundary scattering throughout the temperature range and only a small effect at very low temperatures was found on the calculated $\kappa_{\ell}$.

It is important to note how the present case is very different from our previous work on SiC.\cite{katre2017exceptionally,dongre2018resonant} In the present work, we observe a significant contribution of both EPS and PDPS to \kapa of a highly-doped system; whereas in our previous work we observed that PDPS was sufficient on its own to correctly predict the \kapa of B-doped cubic SiC owing to the resonant phonon scattering that boron causes. \cite{katre2017exceptionally,dongre2018resonant} The resonant scattering was at least one to two orders of magnitude higher than that caused by other defects and resulted in a drastic reduction (approximately two orders of magnitude at room temperature) in the thermal conductivity even at relatively modest defect concentrations ($\approx 10^{20}$~cm$^{-3}$). Boron does not cause resonant scattering in Si and therefore the contribution of both EPS and PDPS are comparable. 

\section{\label{sec:concl}Conclusions}
We have calculated the \textit{ab-initio} lattice thermal conductivity of highly P- and B-doped Si considering phonon scattering caused by three phonon processes, isotopes, point defects, and electrons. We illustrate that at low doping concentrations EPS causes a higher reduction in \kapa as compared to PDPS owing to a near constant behavior of EPS rates at low frequencies. At low concentrations EPS alone is sufficient to reproduce the absolute reduction in $\kappa_{\ell}$. However, at high doping concentrations it fails to reproduce the absolute values and the trend of \kapa reduction. PDPS has a substantial contribution to \kapa reduction at high doping concentrations and neither EPS nor PDPS is sufficient on its own to reproduce the experimental \kapa values. Only when the effect of all the scattering mechanisms are considered together, we get a good agreement with the experimental values across the temperature range. At a doping concentration of $10^{21}~\mathrm{cm}^3$, we observe almost 90\% reduction in the room temperature thermal conductivity of B-doped Si as compared to the bulk, whereas, 80\% reduction in case of P doping is found. This is mainly because of the higher PDPS rates of the B defects than those of P defects. Neither EPS nor PDPS can be captured by conventional parameterized models. Together with our previous works on doping diamond, SiC, GaN and GaAs\cite{ katcho2014effect, katre2017exceptionally, katre2018phonon, kundu2019effect}, we show that, at temperatures above 300 K, PDPS is the most dominant scattering mechanism in highly B- and P-doped Si and cannot be neglected.

\section{\label{sec:data} Data Availability}
The data that support the findings of this study are available from the corresponding author upon reasonable request.
  
\section{\label{sec:ack} Acknowledgments}
The authors acknowledge support from the European Union's Horizon 2020
Research and Innovation Action under Grant No. 645776 (ALMA), the French (ANR) and Austrian (FWF) Science Funds under project CODIS (ANR-17-CE08-0044-01 and FWF-I-3576-N36), and the Natural Science Foundation of China under Grant No. 11704258. We also thank the Vienna Scientific Cluster for providing the computational facilities (project numbers 645776: ALMA and 1523306: CODIS). 

\section{\label{sec:auth} Author Contributions}
G.K.H.M. and B.D. conceptualized the study. J.C. wrote the phonon-defect scattering code. B.D. carried out the phonon-defect scattering calculations and prepared the manuscript. S.W. and J.M. carried out the electron phonon scattering calculations. All authors discussed the results and contributed to the writing of the article.

\section{\label{sec:add} ADDITIONAL INFORMATION}
\textbf{Competing interests}: The authors declare no competing interests.

\bibliographystyle{apsrev4-1}
\bibliography{references.bib}

\end{document}